\documentclass[12pt]{iopart}
\pdfoutput=1 
\usepackage[T1]{fontenc} 
\usepackage{mathrsfs, amssymb,amsfonts,amsthm,graphicx, epsf, yfonts, array}
\expandafter\let\csname equation*\endcsname\relax
\expandafter\let\csname endequation*\endcsname\relax
\usepackage{amsmath}
\usepackage{xcolor}
\usepackage{cite}
\usepackage{enumitem}
\usepackage[hyperfootnotes=true,hyperindex,breaklinks, colorlinks=true, linkcolor=blue, citecolor=blue, urlcolor=blue]{hyperref}
\usepackage{color}
\usepackage{orcidlink}
\DeclareMathAlphabet{\mathpzc}{OT1}{pzc}{m}{it}
\def\Z#1{_{\lower2pt\hbox{$\scriptstyle#1$}}}

\def\goesas{\mathop{\sim}\limits}
\def\lsim{\mathop{\hbox{${\lower3.8pt\hbox{$<$}}\atop{\raise0.2pt\hbox{$\sim$}}$}}}
\def\vZ{v\Z{Z}} \def\vD{v\Z{D}} \def\vK{v\Z{K}} \def\vS{v\Z{S}} \def\vSZ{v\Z{SZ}} 
\def\dd{\mathop{\text{d}\!}}
\def\gphiphi{e^{-2\Phi/c^2}r^2 - c^2A^2e^{2\Phi/c^2}}
\def\rootgphiphi{\sqrt{g_{\phi\phi}}}
\def\AA{{\mathpzc A}} \def\BB{{\mathpzc B}} \def\CC{{\mathpzc C}}
\def\aa{{\mathpzc a}} \def\bb{{\mathpzc b}}
\def\tann{{\mathop{\text{\,tann}\!}}}
\def\be{b_\epsilon}

\begin{document}
\title{Exact solutions for differentially rotating galaxies in general relativity}

\author{Marco Galoppo\orcidlink{ 0000-0003-2783-3603} and David L.~Wiltshire\footnote{Author to whom any correspondence should be addressed.}\orcidlink{0000-0003-1992-6682}}

\address{School of Physical \& Chemical Sciences, University of Canterbury, \\ Private Bag 4800, Christchurch 8140, New Zealand}
\eads{\mailto{Marco.Galoppo@canterbury.ac.nz}, \mailto{David.Wiltshire@canterbury.ac.nz}}
\begin{abstract}
Two classes of stationary axisymmetric solutions of Einstein's equations for isolated differentially rotating matter sources are presented. The asymptotic regime is extracted, with attention to quasilocal gravitational energy, shear and angular momentum related by the Raychaudhuri equation. At outer boundaries the quasilocal angular momentum and energy densities vanish, defining novel geometric structures -- the vortex surface and the rotosurface. Solutions with a rotosurface present a new notion of asymptotic flatness: vorticity, shear and Kretschmann scalars vanish at a finite radial distance from the symmetry centre. Synthetic rotation curves for the Milky Way are presented, fit to GAIA-DR3 data, and potential astrophysical signatures are suggested. Whether or not abundant collisionless dark matter exists, the new solutions very strongly support suggestions that the phenomenology of galactic rotation curves be fundamentally reconsidered, for consistency with general relativity.
\end{abstract}

\maketitle

An exact solution of Einstein's equations for an isolated galaxy sourced by a realistic distribution of stars, treated as an effective pressureless fluid---{\em dust}---is a decades-old open problem that has confounded mathematical relativists for decades. In this Letter we present a new solution to this problem, which unlike previous unsuccessful attempts incorporates the essential physical feature that radial distributions of stars in galaxies rotate by varying amounts. This statement will be precisely defined -- but, roughly, each ring of stars in an axially symmetric distribution rotates at slightly different speeds to its neighbours. A consequence of our result is that the amount of collisionless dark matter that is conventionally inferred needs to be fundamentally revisited, with profound implications for astrophysics and cosmology.

This problem has been the subject of controversy on account of researchers artificially treating galaxies as rigidly rotating objects \cite{Cooperstock_2007,Cooperstock_2008,Balasin_2008}, which is not only physically unrealistic but is also compounded by mathematical problems \cite{Vogt_2005,Neill_2011}. Recently, it was shown that the physical viability of general relativistic galaxy models improved with the introduction of realistic differential rotation \cite{Astesiano_2022a}. {Indeed, we note that differential rotation is a necessary condition for self-gravitating disc systems to achieve dynamical stability in Newtonian dynamics \cite{Toomre_1964, Ostriker_1973, Papaloizou_1984, Pichon_1997, Bertin_2014}.} Remarkably, we will find that the mathematical pathologies can be sidestepped within this wider class of models, by refining the consistent treatment of taking Newtonian limits. In particular, conventionally researchers have na\"ively applied nonrelativistic limiting procedures to the gravitational metric before attempting to include nonlinear terms. By contrast, we will include all essential nonlinearities for stationary axisymmetric dust spacetimes before considering the low--energy limit.

The nonlinearity of general relativity arises from the self--interaction of matter and geometry via Einstein's equations. In place of additive gravitational potentials on a fixed background, the time--averaged motion of matter sources defines regional backgrounds with their own quasilocal energy and angular momentum content \cite{Szabados_2009,Wiltshire_2007,Wiltshire_2011}. {In galaxies, each star is considered to move within an effective fluid element whose internal energy includes not only the rest mass energy of the stars and gas, but also their interactions.} Understanding the hierarchy of regional scales, or the fitting of one regional geometry into another, is an important foundational question in cosmology \cite{Ellis_1984}. Conventionally one assumes the existence of a global asymptotic Minkowski background before applying nonrelativistic limits. Mass and angular momentum are then defined by ideal asymptotic Killing vectors, which obscures their essential quasilocal origin in general relativity. The manner in which quasilocal energy and angular momentum approach asymptotic values in a nonempty universe is little appreciated. It may nonetheless be key to understanding the biggest open problems in physical cosmology.

In this Letter we derive new solutions with self-consistent coupling between quasilocal energy and angular momentum, which can be applied in {both the low-energy and ultrarelativistic limits} to systems at different scales in the fitting problem. Here our primary interest is rotating galaxies.
However, our approach will likely further inform other attempts to model the dynamical properties of coarse-grained, cosmological structures in general relativity \cite{Giani_2022,Hills_2022,Okeeffe_2023,Giani_2024,Seifert_2024}. {The new exact solutions pave the way not only as a background for linear stability analyses incorporating relativistic aspects of angular momentum in general relativity, but ultimately for the more complex issue of the nonlinear stability of differentially rotating disc galaxies in general relativity.}

Let us consider stationary axisymmetric solutions of the Einstein equations with a dust source. Our solutions will thus apply to numerous astrophysical systems including stellar systems, galaxies and putative dark matter halos on larger scales. As applied to galaxies, assuming stationarity---i.e., the presence of a timelike Killing vector, $\partial/\partial t$---means the solutions apply to short time scales relative to those involving galaxy formation and evolution, denoted $t_{\text{evo}}$. Likewise, imposing axisymmetry---i.e., an axial Killing vector, $\partial/\partial\phi$---means that our solutions then apply to an effective dust fluid for time-averaged oscillatory motion, $t_{\text{osc}}$, of individual stars and gas above and below the galactic plane. At the present epoch $t_{\text{osc}}\goesas10\,$Myr and $t_{\text{evo}}\goesas1\,$Gyr. Thus the effective dust fluid applies on time scales, $10^7\lsim t\lsim10^9\,$yr for nearby galaxies. {Furthermore, since gravitational degrees of freedom are included in the effective fluid elements, their energy density does not need to remain strictly positive everywhere although all particles averaged over must satisfy the positive energy condition.}

The spacetime metrics we consider can thus be written in the Lewis--Papapetrou--Weyl form, viz.
\begin{equation}
 \label{eq:metric}
 \dd s^2 =-c^2 e^{2\Phi(r, z)/c^2}(\dd t+A(r, z) \dd\phi)^2 +e^{-2\Phi(r, z)/c^2}\left[r^2 \dd\phi^2 +e^{2k(r, z)/c^2}(\dd r^2+\dd z^2)\right]\, ,
\end{equation}
where $r$ is a radial coordinate, $z$ is an axial coordinate, $\Phi(r,z)$ is related to the conventional Newtonian potential, $A(r,z)$ is related to frame-dragging, and $k(r,z)$ is a conformal factor on the 2--dimensional space of orbits of the isometry group generated by the Killing vectors.

The energy-momentum tensor takes the form
\begin{equation}
\label{eq:EMT}
 T^{\mu\nu}=c^2\rho(r, z) \, U^{\mu}U^{\nu}\, ,
\end{equation}
where $\rho(r,z)$ is the effective density of coarse-grained fluid elements, each with a 4--velocity $U^\mu$, given by
\begin{equation}
\label{eq:U}
 U^{\mu}\partial_\mu=(-H)^{-1/2}\left(\partial_t +\Omega \, \partial_{\phi}\right)\, ,
\end{equation}
where $d\phi/dt=\Omega(r, z)$ uniquely defines the angular speed of rotation at any point, and $H(r, z)$ is a normalization factor. Since $U^{\mu}U_{\mu}=-c^2$, it follows that
\begin{equation}
\label{eq:Hnorm}
  H=-e^{2\Phi/c^2}(1+A \, \Omega)^2+e^{-2\Phi/c^2}r^2 \, \Omega^2/c^2\, .
\end{equation}

Einstein's equations are solved by a hierarchy of quadratures \cite{Geroch_1971,Geroch_1972,Winicour_1975,Stephani_2003}. One finds that $H=H(\eta)$ and $\Omega=\Omega(\eta)$, where $\eta(r,z)$ is the angular momentum of the dust elements measured by Zero Angular Momentum Observers (ZAMOs), and
\begin{equation}
\label{eq:OmH}
  \dd\Omega = c^2\dd H/({2\eta})\,.
\end{equation}
In view of \eqref{eq:OmH} we call $H$ the {\em differential rotation state function}, since a choice of $H$ fixes $\Omega$. The norms of the two Killing vectors, $\xi^\mu\partial_\mu = \partial_t$ and $\zeta^\mu\partial_\mu = \partial_\phi$, and their inner product are then respectively given by
\begin{align}
 & \xi^\mu \xi_\mu = - c^2e^{2\Phi/c^2} = \frac{\left(cH-\eta \, \Omega/c\right)^2 - \left(r\,\Omega\right)^2}{H} \, , \label{eq:XiXi}\\
 & \zeta^\mu \zeta_\mu = r^2 e^{-2\Phi/c^2} -c^2 A^2 e^{2\Phi/c^2} = \frac{r^2-(\eta/c)^2}{(-H)}\,, \label{eq:ZetaZeta} \\ & \xi^\mu \zeta_\mu = -c^2 A e^{2\Phi/c^2} = \eta - \frac{(\eta/c)^2-r^2}{H}\,\Omega \label{eq:XiZeta} \,.
\end{align}
The remaining Einstein equations yield
\begin{align}
 & \Xi_{,r} = \frac{1}{2r}\left[g_{tt,z}g_{\phi\phi,z}-g_{tt,r}g_{\phi\phi,r}-(g_{t\phi,z})^2+(g_{t\phi,r})^2\right]\,,\nonumber\\
 & \Xi_{,z} = \frac1{2r}\left[2g_{t\phi,r} g_{t\phi,z}+g_{tt,r}g_{\phi\phi,z}-g_{tt,z}g_{\phi\phi,r}\right]\,,\label{eq:Xirz}
\end{align}
where $\Xi(r,z) = [\Phi(r,z) - k(r,z)]/c^2$.
{Following Geroch \cite{Geroch_1971,Geroch_1972} and Winicour \cite{Winicour_1975}, (as reviewed in Ref.\ \cite{Stephani_2003}), we finally require} $\Xi_{,rz}=\Xi_{,zr}$ to find that the entire class of solutions is fully determined by a choice of $H$ and a solution of the homogeneous Grad--Shafranov equation \cite{Astesiano_2022a,Ruggiero_2024,Re_2024}
\begin{equation}
 \label{eq:Grad-Shafranov}
 \Psi_{,rr}-\frac{1}{r}\Psi_{,r} + \Psi_{,zz} = 0,
\end{equation}
where
\begin{equation}
 \label{eq:IntegralRelation}
 \Psi = \eta + c^2\frac{r^2}{2} \int \frac{H'}{H} \frac{d\eta}{\eta} - \frac{1}{2} \int \frac{H'}{H}\eta\, d\eta\, .
\end{equation}

Equation \eqref{eq:IntegralRelation} is an integrability condition for the dust geodesic equations in the absence of pressure in the effective fluid {\cite{Geroch_1971,Geroch_1972, Winicour_1975, Stephani_2003}}. Furthermore, to model galaxies, we are interested in solutions with reflection symmetry with respect to the $z = 0$ plane. The solution to \eqref{eq:Grad-Shafranov} is then given by
\begin{align}
 \label{eq:GeneralSolution}
 \Psi(r,z) = \,\CC_0 \;+\; &r\sum_{n=0}^\infty \AA_n I_1(\aa_n r)\cos(\aa_n z) + r\sum_{m=0}^\infty \BB_m K_1(\bb_m r)\cos(\bb_m z), 
\end{align}
where $I_1$ and $K_1$ are the modified Bessel functions of the first and second kind, and the constants $\aa_n, \bb_m$, $\CC_0$, $\AA_n$ and $\BB_m$ {must be taken to be real numbers. The first series is regular on the axis of rotation, whilst the second series is regular at spatial infinity.}

The trace of the Einstein equations gives
\begin{equation}
 \label{eq:density}
 8\pi G\rho = \frac{\eta_{,r}^2+\eta_{,z}^2}{4\eta^2 r^2}\left[\eta^2\Delta^2(\eta) - c^4r^4\ell(\eta)^2\right]e^{2\Xi}\!, 
\end{equation}
where $\ell(\eta) = H'/H$ and $\Delta(\eta)=2-\eta\,\ell(\eta)$, with $H'=dH/d\eta$,
while the Raychaudhuri equation reduces to
\begin{equation}\label{eq:Raychaudhuri}
 \dot\Theta\equiv U^\nu \nabla_\nu\Theta=-\frac{1}3\Theta^2 +\omega^2-\sigma^2 -R_{\mu\nu}U^\mu U^\nu ,
\end{equation}
where $\Theta=\nabla_\mu U^\mu$, $\omega^2=\omega^{\mu\nu} \omega_{\mu\nu}$ and $\sigma^2= \sigma^{\mu\nu} \sigma_{\mu\nu} $ are the expansion, vorticity and shear scalars respectively. Here $\omega_{\mu\nu}= U_{[\mu;\nu]}$, $\sigma_{\mu\nu}= U_{(\mu;\nu)}-\frac13 h_{\mu\nu} \Theta$, $h_{\mu\nu}= g_{\mu\nu}+c^{-2}U_\mu U_\nu $. The dust is non-expanding, $\Theta=0$, $\dot\Theta=0$, and $R_{\mu\nu}U^\mu U^\nu=4\pi\,G\rho$, so that \eqref{eq:Raychaudhuri} reduces to a balance condition
\begin{equation}\label{eq:Raybalance}
 \omega^2-\sigma^2 = 4\pi G\rho\,.
\end{equation}
Furthermore, once effective pressure, $P$, is included \cite{Galoppo_2024b}, the r.h.s.\ of \eqref{eq:Raybalance} becomes $4\pi G(\rho+3P/c^2)$, also for strong gravitational fields.
Significantly, since the fluid element energy density contains quasilocal kinetic energy, $\rho$ may be negative when shear dominates over vorticity. Indeed,
\begin{equation}\label{eq:shear}
 \sigma^2 = \frac{e^{2\Xi}c^4r^2(\eta_{,r}^2+\eta_{,z}^2)\,\ell^2(\eta)}{8\eta^2},
\end{equation}
which combined with \eqref{eq:density} yields
\begin{equation}
 \label{eq:vorticity}
 \omega^2 = \frac{e^{2\Xi}(\eta_{,r}^2+\eta_{,z}^2)\,\Delta^2(\eta)}{8r^2}\,.
\end{equation}

Physical measurements are defined by relevant classes of observers. We first introduce the ZAMO coframe \cite{Bardeen_1972,Crosta_2020,Beordo_2024}, {which is completely determined by the requirement of zero angular momentum and by the 1+3 splitting of spacetime, as}
\begin{align}
{\mathbf \omega}\Z Z=\Bigl(\frac{r\dd t}{\rootgphiphi}\,,\rootgphiphi &\left(\dd \phi \,-\,\chi \dd t\right)\,,e^{(k-\Phi)/c^2} \dd r\,,e^{(k-\Phi)/c^2} \dd z\Bigr)\,, \label{eq:Zcoframe}
\end{align}
$g_{\phi\phi}=\gphiphi$, and $\chi= - g_{t\phi}/g_{\phi\phi}$ with $g_{t\phi}= - c^2\,A\,e^{2\Phi/c^2}$.
The dual ZAMO tetrad frame is then
\begin{align}
{\mathbf e}\Z Z=\Bigl(\frac1{r}{\rootgphiphi}&\left(\partial_t\,+\,\chi\partial_\phi\right)\,,\frac{1}{\rootgphiphi}\partial_\phi\,, e^{-(k-\Phi)/c^2} \partial_r\,,e^{-(k-\Phi)/c^2} \partial_z\Bigr)\,.\label{eq:Zframe}
\end{align}
Starting from the formula for the dust elements' velocity measured by ZAMOs, $\vZ =({{{\mathbf \omega}\Z Z}^1}\cdot{\mathbf U})/({{{\mathbf \omega}\Z Z}^0}\cdot{\mathbf U})$, we find
\begin{equation}
 \label{eq:eta}
 \eta(r,z) = r\, \vZ (r,z) \, .
\end{equation}
Furthermore, let us define an effective Lorentz factor $\gamma\Z Z:= {{\mathbf \omega}\Z Z}^0\cdot{\mathbf U}$, 
so that 
\begin{equation}
 \label{eq:vZ}
  -H \,\gamma_\mathrm{Z}^2 \,\vZ=r\,(\Omega-\chi)\,.
\end{equation}

The second congruence of relevance are ideal Stationary Observers (SOs), {i.e., observers who are nonrotating with respect to the given coordinates,}
with coframe \cite{Costa_2023}
\begin{align}
{\mathbf \omega}\Z S=\Bigl(e^{\Phi/c^2}&\left(\dd t \,+\,A \dd \phi\right)\,, r\,e^{\Phi/c^2} \dd \phi\,,e^{(k-\Phi)/c^2} \dd r\,,e^{(k-\Phi)/c^2} \dd z\Bigr)\,.\label{Scoframe}
\end{align}
The dual SO tetrad frame is then
\begin{align}
{\mathbf e}\Z S=\Bigl(e^{-\Phi/c^2}&\partial_t\,,\frac{e^{-\Phi/c^2}}{r}\left(\partial_\phi-A\,\partial_t\right)\,, e^{-(k-\Phi)/c^2} \partial_r\,,e^{-(k-\Phi)/c^2} \partial_z\Bigr)\,.\label{Sframe}
\end{align}
By analogy to the ZAMO case for SOs we define, $\vS/c =({{\mathbf \omega}\Z S^1}\cdot{\mathbf U})/({{\mathbf \omega}\Z S^0}\cdot{\mathbf U})$ and $\gamma\Z S:= {{\mathbf \omega}\Z S}^0\cdot{\mathbf U}$, so that
\begin{equation}
 \label{eq:eatthat}
 \vS (r,z) =\frac{e^{\Phi/c^2}r\Omega}{\gamma_\mathrm{S}{\sqrt{-H}}}\,.
\end{equation}
We can define $\eta\Z S=r\vS$ by analogy to \eqref{eq:eta}. However, it plays no particular role here. Equivalently, {the relative velocity between the two frames is given by}
\begin{equation}
\label{eq:diff}
 \vSZ=\frac{{{\mathbf \omega}\Z S^1}\cdot{{\mathbf e}\Z Z^0}}{{{\mathbf \omega}\Z S^0}\cdot{{\mathbf e}\Z Z^0}} =\frac{g_{\phi\phi}}{r^2}r\chi=\frac{g_{t\phi}}{r} , 
\end{equation}
so that
\begin{equation}
 \label{eq:vStovZ}
  e^{\Phi/c^2}\gamma\Z S \vS=-H \,\gamma_\mathrm{Z}^2 \vZ + \frac{r^2}{g_{\phi\phi}}\vSZ\,.
\end{equation}
Eqs.~\eqref{eq:diff},\eqref{eq:vStovZ} show that ZAMOs and SOs will measure different dust velocities at any spacetime event, with a difference directly related to the frame-dragging contribution.

In the present framework it is clear how observations have been misapplied to theoretical observables in the past. In particular, several analyses \cite{Cooperstock_2007,Cooperstock_2008,Balasin_2008} misapply $\vZ$ to the rotation curves of distant galaxies, when $\vS$ given by \eqref{eq:eatthat} is the relevant velocity, since {
\begin{equation}
  \label{eq:vSToRedshift}
  \frac{\vS}{c} = \frac{e^{+\Phi/c^2}r\Omega/c}{\sqrt{H^2+\left(e^{+\Phi/c^2}r\Omega/c\right)^2}}\simeq\frac{r\Omega}{c}+\dots
\end{equation}}
coinciding, at leading order, with the widely used special relativistic interpretation of the redshift, {$\text{z}$, as applied to this class of spacetimes in the low velocity limit \cite{Re_2024}, namely 
\begin{equation}
\label{eq:redshift2}
  1+\text{z} \simeq 1+\frac{\vK}{c}\cos\theta+c^{-2}\left(\frac{\vK^2}{2}-\Phi+2\vK\vD\right)+ ...\, ,
\end{equation}
where $\theta$ is the angle between the photon trajectory and the direction $-\hat{\phi}$, identified via $\partial_\phi$.} Moreover, to complete our identification of physically relevant velocities in the general case, we supplemented the velocities $\vZ$ and $\vS$ by the kinetic and dragging velocities \cite{Astesiano_2022b, Re_2024}
\begin{align}
  & \vK := r\,\Omega \, , \nonumber\\
  & \vD := r\, \chi \, ,\label{eq:RelVelocity} 
\end{align}
respectively. We note that whilst $\vS$ and $\vK$ coincide at low velocities, in general their differences can be physically important.

Let us now derive the functional form applicable to systems such as galaxies, with subrelativistic local relative speeds, $v\ll c$, weak pseudo-Newtonian potentials, $\Phi\goesas v^2/c^2$, and nonrelativistic frame-dragging \footnote{{In the low-energy limit of the class of spacetimes we consider it necessarily follows that $\vZ \goesas\vD\goesas \vS \goesas \vK \goesas v \ll c$, as can be seen from Eqs. \eqref{eq:vZ}, \eqref{eq:diff} and \eqref{eq:vSToRedshift}. See Ref.\ \cite{Re_2024} for a more in-depth discussion of taking such a limit in the $(\eta, H)$ class of solutions.}}.

From these conditions, we find $A = r\,\vD /c^2\left[1+\mathcal{O}({v^3/c^3})\right]$. Thus by \eqref{eq:Hnorm}, $ H = -1 + \mathcal{O}({v^2/c^2})$. Since $H = H(\eta)$ {from \eqref{eq:OmH}}, using \eqref{eq:eta} 
we find
\begin{equation}
\label{eq:Heta}
  H(\eta) = -1 + \epsilon\, \eta^2/(\be c)^2 + \mathcal{O}(v^3/c^3) \,, 
\end{equation}
where $\epsilon\in\{-1,0,1\}$ and $\be$ are constant lengths which can be identified as impact parameters.
We stress that the functional form of $H(\eta)$ in \eqref{eq:Heta} is the only possible choice that allows a dust system to be consistently considered in a low-energy regime.

{Having established the self-consistent form \eqref{eq:Heta} of $H(\eta)$ in the low energy limit, we will now adopt the global ansatz, $H(\eta) = -1 + \epsilon\, \eta^2/(\be c)^2$, to determine new exact solutions of Einstein's equations. These solutions can now be applied even at high energies and in the ultrarelativistic limit as long as the underlying hypotheses remain valid. Substituting the ansatz for $H(\eta)$}
into \eqref{eq:IntegralRelation} we obtain
\begin{equation}
  \label{eq:EtaGW}
  \eta_\epsilon(r,z) =\be c\tann\left({\frac{\be\,\Psi(r,z)}{\be^2c-\epsilon\,r^2c}}\right)\,,
\end{equation}
where
\begin{equation}
  \tann\,(x)=
\begin{cases} \tanh(x), &\epsilon=+1\\ x,&\epsilon=0\\ \tan(x),&\epsilon=-1\\
\end{cases}.
\end{equation}

Moreover, by \eqref{eq:OmH} it follows that
\begin{equation}
  \label{eq:OmegaGW}
  \Omega_\epsilon = \Omega_0 + \epsilon \frac{\eta_\epsilon}{\be^2}.
\end{equation}
Furthermore, to ensure zero rotation on the symmetry axis we choose $\Omega_0 = 0$. We thus find 
\begin{equation}
  \label{eq:vKGW}
  \vK = \epsilon\frac{rc}{\be}\,\tann\left({\frac{\be\,\Psi(r,z)}{\be^2c-\epsilon\,r^2c}}\right)\,,
\end{equation}
and 
\begin{equation}
  \label{eq:vDGW}
  \vD = c\left(\frac{r}{\be} \epsilon-\frac{\be}{r}\right)\,\tann\left({\frac{\be\,\Psi(r,z)}{\be^2c-\epsilon\,r^2c}}\right).
\end{equation}

The presence of differential rotation is crucial mathematically since the term $\epsilon r^2 c$ in \eqref{eq:EtaGW} regularizes potential divergences at infinity in \eqref{eq:GeneralSolution}. In particular, we can choose solutions to \eqref{eq:Grad-Shafranov} which are regular about the $z$-axis and diverge as $\Psi\goesas r^{\alpha}$, $\alpha < 2$ as $r\to\infty$. All these cases are phenomenologically interesting as the sub-quadratic divergences in the modified Bessel function series $I_1(\aa_n r)$ in \eqref{eq:GeneralSolution}, are thereby regularized. This option was unavailable in the rigidly rotating Balasin-Grumiller (BG) model 
\cite{Balasin_2008}\footnote{The same equations are solved in a different context by Neugebauer and Meinel \cite{Neugebauer_1993,Neugebauer_1994,Neugebauer_1995} {with a first generalisation to differential rotation by Ansorg and Meinel in \cite{Ansorg_2000}}.} where only the modified Bessel function series $K_1(\bb_n r)$ in \eqref{eq:GeneralSolution} was chosen, being regular as $r\to\infty$ but singular on the $z$-axis. Thus, we find solutions that display self-consistency in the low-velocity regime. 

By \eqref{eq:density} and \eqref{eq:Heta} in each limit the density of effective fluid elements is 
\begin{equation}
\label{eq:DensityGW}
  8\pi G\rho_\epsilon = \frac{\eta_{\epsilon,r}^2+\eta_{\epsilon,z}^2}{2r^2}\left[\frac{\be^4-\epsilon^2 r^4}{\left(\be^2-\eta_\epsilon^2/c^2\right)^2}\right]e^{2\Xi_\epsilon}\, .
\end{equation}
{Thus we identify three families of solutions that exhibit distinct physical behaviours, depending on the choice of the parameter $\epsilon$. For $\epsilon = 0$ we recover the unphysical rigidly rotating class\footnote{{Equation \eqref{eq:Raybalance} shows that the density must be identified solely with the vorticity of the fluid. Therefore, in the resulting spacetime, the gravitational field would be determined exclusively by the magnetic part of the Weyl tensor, which does not possess a Newtonian analogue \cite{Wainwright_Ellis_1997}. Consequently, such solutions cannot be considered to be physically viable.}}. 

The differentially rotating $\epsilon = \pm 1$ solutions can instead be understood in terms of a competition between the matter density and angular momentum density in determining the local dragging of one shell of matter relative to shells a small radial distance away. Since vorticity acts outward, in both cases in the interior where $\rho_\pm>0$, local frame dragging acts outward and the angular momentum density peaks at larger radii than the matter density. Thereafter the dragging of shells will act radially inward. Eventually we encounter surfaces at which, $\rho_\pm=0$, beyond which formally $\rho_\pm<0$, signaling a breakdown of the approximations used. In the exterior other forms of quasilocal gravitational energy become important for the effective fluid elements. 

The distinction between the $\epsilon=-1$ and $\epsilon=+1$ spacetimes turns out to be determined by the steepness of the gradients of the density and angular momentum density, while maintaining the equilibrium relation \eqref{eq:Raychaudhuri}. Mathematically, two possibilities arise for the bounding surfaces where $\rho_\pm=0$, located at $r = b_\pm$. For $\epsilon = -1$ the angular momentum density per unit mass is locally monotonically decreasing as $r\to b_-$, whilst for $\epsilon = +1$ it is locally monotonically increasing on each side of an apparent surface of discontinuity at $r=b_+$. We denote $r=b_-$ as the {\em vortex surface}, and $r=b_+$ as the {\em rotosurface} respectively.}

{For the $\epsilon=-1$ spacetimes, as $r\to b_-$, $\vZ$ and $\vS$ are antiparallel.} The vortex surface is found at $r=b_-\gg L$, where $L$ is a typical scale length of the coarse-grained dust. E.g., for the Milky Way, $L \goesas 50\,$kpc. By \eqref{eq:ZetaZeta}, \eqref{eq:EtaGW}, on the vortex surface
\begin{equation}
  g_{\phi\phi} = b_-^2\left[\left(1-\frac{1}{2}\frac{\Psi(r,z)^2}{c^2b_-^2}\right) + \mathcal{O}(v^4/c^4)\right] \, ,
\end{equation}
and other metric components are also regular, all relative velocities, $v$, being subrelativistic.

{For the $\epsilon=+1$ spacetimes, as $r\to b_+$, $\vZ$ and $\vS$ are parallel.} The rotosurface, $r=b_+$, has the following novel geometric properties. In the limit $r\to b_+$ at finite $z$, even though the Killing vector norms diverge, the vorticity, shear, density, Ricci and Kretschmann scalars converge to zero: 
\begin{equation}
\omega^2=\sigma^2= R_{\mu\nu\alpha\beta}R^{\mu\nu\alpha\beta}\simeq \frac{e^{2\Xi}\Psi^2}{8|b_+-r|^4}\to 0, 
\end{equation}
since $\Psi$ is finite and $e^{2\Xi}\goesas e^{-\Psi^2/(b_+c^2|b_+-r|)}$ as $r\to b_+$, {as follows from a direct expansion of the system of equations \eqref{eq:Xirz} in powers of $|b_+-r|^{-1}$.}

The regular rotosurface is a new mathematical object, whose topology, geometry, fundamental physics and cosmological implications remain to be fully explored. Analogously to an acceleration horizon for uniform linear acceleration in Minkowski space, it is a limiting rotating surface where a fictitious observer would need to rotate with $v\to c$ to have zero angular momentum with respect to the vanishing local dust. However, it is a timelike surface. Indeed, it can provide a precise mathematical definition of one type of finite infinity surface \cite{Wiltshire_2007,Wiltshire_CEP_2008,Wiltshire_2014}. Since the rotosurface is regular but timelike it may provide a novel mathematical arena for understanding {asymptotia}.

{For both the $\epsilon=\pm1$ classes, the quasilocal energy density of the effective fluid is negative in the exterior. However, its magnitude remains small and the addition of other terms yields an overall total positive energy density. In the exterior} other quasilocal energy sources will dominate: the effective thermal pressure of galaxies in clusters; and the kinetic energy of expansion in a void dominated universe \cite{Williams_2024}, or alternatively a cosmological constant, $\Lambda$. The fact that the quasilocal energy of the isolated system is negative beyond the bounding surface is consistent with: (i) positive spatial curvature energy and binding energy being negative relative to the background; (ii) a small kinematical backreaction, $\Omega_{\cal Q}$, of virialized structures that is negative in the Buchert averaging scheme \cite{Buchert_2000,Buchert_2020}. For a cosmological constant, $\rho_\Lambda=3 H_0^2\Omega_{\Lambda0}/(8\pi G)=88.3\;\mathrm{M}_\odot\;$kpc$^{-3}$, $\rho_\Lambda+\rho>0$. For a typical {massive disc galaxy} solution with inner baryonic galactic density $\rho_{\text{baryon}}\sim10^9$--$10^{11}\;\mathrm{M}_\odot\;$kpc$^{-3}$, $\rho_\Lambda$ exceeds the maximum value of $-\rho_\pm$ by a factor $\sim1.5$--$1.8$.

\begin{figure}[htb!]
  \centering
  \includegraphics[width = .95\textwidth]{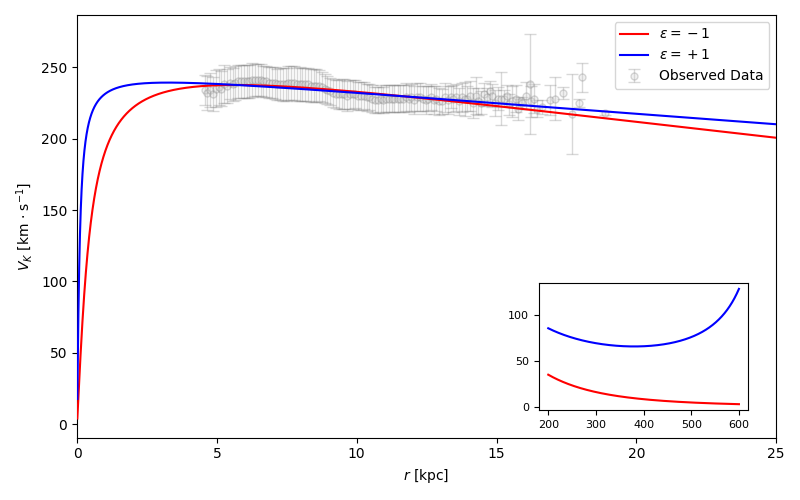}
  \caption{Rotation velocity curves for both $\epsilon=-1$ and $\epsilon=+1$ spacetimes fit to the GAIA-DR3 data sample ``ALL'' of Beordo {\em et al}.~\cite{Beordo_2024} (superposed). Insets show the approach $r\to b_-$ and $r\to b_+$.}
  \label{fig:RotationCurve}
\end{figure}

\begin{figure}[htb!]
  \centering
  \includegraphics[width = .95\textwidth]{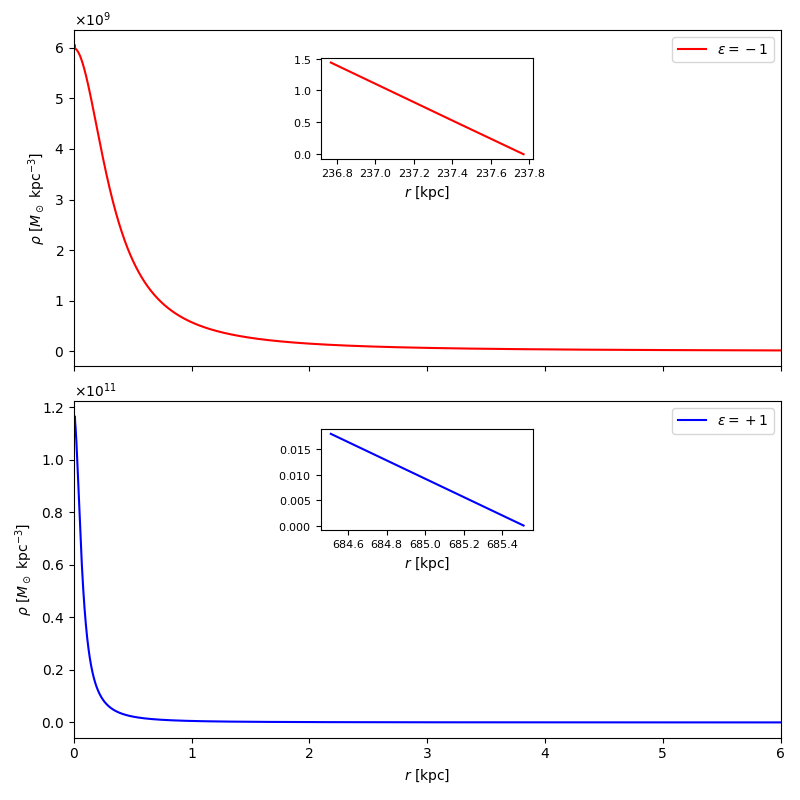}
  \caption{Radial effective dust density distribution for both $\epsilon=-1$ and $\epsilon=+1$ spacetimes obtained via the fit to GAIA-DR3 data. Insets show the approach $r\to b_-$ and $r\to b_+$.}
  \label{fig:Densities}
\end{figure}

\begin{figure}[htb!]
  \centering
  \includegraphics[width = .95\textwidth]{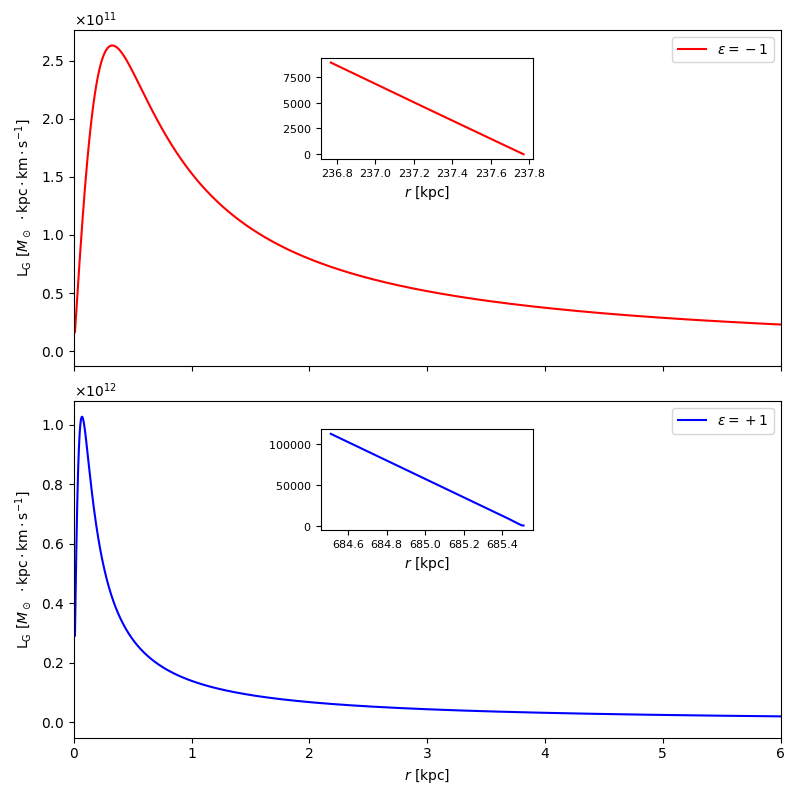}
  \caption{Angular momentum density distribution measured by the stationary observer, i.e., $\text{L}_{\text{G}}:=r\rho\vK$ for both $\epsilon=-1$ and $\epsilon=+1$ spacetimes obtained via the fit to GAIA-DR3 data. Insets show the approach $r\to b_-$ and $r\to b_+$.} \label{fig:AngularMomentum}
\end{figure}

Since the new solutions successfully exhibit the essential physics of differential rotation they apply naturally to rotationally supported systems, e.g., disc galaxies. Sample exact solutions for a Milky Way-like galaxy fit to GAIA-DR3 data\footnote{{We use the sample ``ALL'' of Beordo {\em et al.}~\cite{Beordo_2024}.} Crosta {\em et al}.~\cite{Crosta_2020,Beordo_2024} considered a number of models, including rigidly rotating general relativistic ones. These were shown to fit the Milky Way rotation curve, as inferred from the proper motion and redshifts of a sample of $7.2\times10^6$ young stars from the GAIA--DR3 survey, homogeneous in phase space, lying within $r<19\,$kpc of the galactic centre and $-1<z<1\,$kpc about the galactic plane. In differentially rotating solutions, we see by \eqref{eq:EtaGW} that for large values of the impact parameter, $b_\epsilon$, almost rigid rotation is obtained. This explains why the $K_1(\bb_n r)$ series alone can be fit to observations close to the galactic plane and far from its centre \cite{Crosta_2020,Beordo_2024}.}
for both the $\epsilon=-1$ and $\epsilon=+1$ spacetimes are shown in figures \ref{fig:RotationCurve}--\ref{fig:AngularMomentum}, which respectively show an observationally inferred rotation curve, the density and radial angular momentum density.


{Both $\epsilon=-1$ and $\epsilon=+1$ solutions fit current GAIA-DR3 data with a comparable reduced $\bar{\chi}^2$ per degree of freedom -- the $\epsilon=+1$ solution fitting somewhat better, with $\bar{\chi}^2_+ = 1.83$ and $\bar{\chi}^2_- =2.17$ respectively. However, they have distinct physical differences which future observational tests can target. Physically the clearest distinction is the stronger gradient of the central density for the $\epsilon=+1$ solution (figure \ref{fig:Densities}), leading to a peak in the radial angular momentum density, {i.e., $\text{L}_{\text{G}}:=r\rho\vK$ in the low-energy regime}, closer to the galactic core (figure \ref{fig:AngularMomentum}). Furthermore the magnitude of the central density is an order of magnitude larger for the $\epsilon=+1$ solution, $\rho_+\sim10^{11}\mathrm{M}_{\odot}\,$kpc$^{-3}$, than for $\epsilon=-1$, $\rho_-\sim 6 \times10^{9}\mathrm{M}_{\odot}\,$kpc$^{-3}$. Standard Newtonian estimates for the central baryonic matter density for the Milky Way give estimates of $\approx 2 \times10^{10}\, \mathrm{M}_\odot\,\mathrm{kpc}^{-3}$ already at $0.3\,$kpc for the disc component, ignoring the bulge contribution \cite{Launhardt_2002,Bland-Hawthorn_2016}. Given uncertainties in standard astronomical modeling we conclude that both $\epsilon=\pm1$ are consistent, though the $\epsilon=+1$ case could more easily accommodate additional particulate dark matter.}

{Finally, a further direct astrophysical consequence of the $\epsilon=+1$} rotosurface is an observable signature similar to that of a shock front. As seen in figure \ref{fig:RotationCurve}, $\vK$, decreases at radii constrained by {\tt H}$\,${\sc i} rotation velocity measurements, but then increases at $r\gtrsim\, 400\,$kpc. As $\vK\to c$ at the rotosurface, the ultra diffuse ions of the intergalactic medium would emit synchrotron radiation as they are swept into the virial exterior of the Milky Way's domain of influence. This source of synchrotron radiation would peak on the plane of the galaxy and might potentially contribute to the radio frequency foregrounds that are modelled empirically and subtracted when determining the spectrum of primordial cosmic microwave background anisotropies \cite{Planck4,Planck11,Planck12}.

A deeper understanding of virialization stands to be opened up by the new exact solutions, not only in the far field limit, but also with the addition of pressure, in the strong field ultrarelatistic limit where quasilocal energy plays a central role in virial relations \cite{uzun_2015}. {Further deepening of understanding of virialization is equally important at low energies: with the addition of pressure the interior solutions \cite{Galoppo_2024b} can be recognized as the first nontrivial physically well-motivated example of Ehler's Newton--Cartan limit \cite{Ehlers_1981,Ehlers_1997,Re_2025}.}

\medskip
\noindent{\bf Acknowledgments}\quad 
This work was supported by Marsden Fund grant M1271 administered by the Royal Society of New Zealand, Te Ap\=arangi. We thank Roy Kerr,\break Federico Re, and Chris Stevens for their physical and mathematical insights that helped us refine our understanding. We thank John Forbes, Christopher Harvey-Hawes, Morag Hills, Emma Johnson, Zachary Lane, Antonia Seifert, Shreyas Tiruvaskar and Michael Williams for useful discussions, and Sergio Cacciatori, Mariateresa Crosta, and Frederic Hessman for correspondence. Finally, we thank an anonymous referee for their detailed analysis of the manuscript, correcting an error which led to significant improvements.
\bibliographystyle{iopart-num}
\bibliography{revolution.bib}

\end{document}